\documentclass[notitlepage,12pt]{article}
 
\usepackage{amsfonts}

\newcommand{\be}{\begin{equation}}
\newcommand{\ee}{\end{equation}}
 
\newcommand{\bea}{\begin{eqnarray}}
\newcommand{\eea}{\end{eqnarray}}

\usepackage{amsfonts} 
\usepackage[dvips]{graphicx}
\usepackage{amssymb} 
\usepackage{amsbsy}
\usepackage{hyperref}
\usepackage{xcolor}
\usepackage{lscape}
\usepackage{booktabs}

\begin{document}


\centerline{\bf
Is there a relationship between Stephen Hawking's worldview}
\centerline{\bf
and his physical disability? On the importance} 
\centerline{\bf
of cognitive diversity}

\hfil







{\footnotesize 

\centerline{
Manuel~Ortega-Rodr\1guez, Hugo~Sol\1s-S\'anchez}

\centerline{
Escuela de F\1sica, 
Universidad de Costa Rica, 11501-2060 San Jos\'e, Costa Rica} }


\hfil





\begin{flushright} 
{\footnotesize 
{\sl 
Despite the adversities---or perhaps, as some have suggested, \\
because of them---he would continue to climb.  \\ Where his feet 
could not go, his mind would soar. 
} \\
J. Gribbin, M. White, {\it Stephen Hawking: A Life in Science}
}
\end{flushright}

\hfil

The following is a comment about the value of thinking from a different perspective. 
It is offered by two active scientists in the field of black hole physics, and it is 
not an intrusion of epistemological relativism into science, but rather an attempt 
to provide insight and encourage constructive reflection. 

It has been a notable issue in the aftermath of Stephen Hawking's death that somebody 
is yet to make a comment on the rich relationship between Stephen Hawking's worldview 
concerning black hole systems and his physical disability. This opportunity would not 
have been missed had the person been a famous artist, for example.

Even though the concepts related to black hole physics have been the creation 
of many scientists, Stephen Hawking has been distinctly influential in the field and has given us and physics a very particular, idiosyncratic and fruitful worldview.

He is of the vision that information is the key concept that must be saved before the star collapses into a black hole. 
It is hard not to notice how the scientist mirrors his living experience into the object of study. His black hole and the related information became embodied.

It is a truism to say that Stephen Hawking did not invent the concept of black hole, 
nor the concept of total gravitational collapse or that of information 
in a physical system. 
However, the way he put these concepts together is absolutely his, and his only.

While many scientists consider the static, end-product black hole and its quantum, 
near-horizon-generated Hawking radiation as the usual picture, 
Stephen Hawking thought more in terms of an evolving, 
collapsing object and how information manages (or fails) to escape before it is too late.

Stephen Hawking was reserved about his disability but grateful 
that his limitations were physical and not mental. 
He actually commented on the benefits of his condition to carry out
his work in theoretical physics, and never lost time regretting the drawbacks [1].

The amyotrophic lateral sclerosis (ALS) is a progressive degenerative disease, 
which causes the death of neurons that control voluntary muscles. 
In 1963, Stephen Hawking received the diagnosis of ALS.  
He became depressed and managed to overcome this depression with the help of family and colleagues [2]. 
Seeing how his body weakened due to a decrease in muscle size, but at the same 
time growing intellectually, he made a research shift from cosmology to black hole singularities, 
but using the then ``unpopular'' Oppenheimer framework [3], 
leading to his 1969 paper [4], where the collapse played a central role in the argument. 
The citations to, and posterior association with, 
Roger Penrose and Yakov Zel'dovich [5], two of the few physicists 
who were citing Oppenheimer articles at the time, evidence such framework choice.
In the gravitational collapse, 
the star collapses by itself reducing its volume and the effects of general relativity become so important that they eventually dominate over all other forces [6].
This can be contrasted with the alternate view that spacetime singularities are 
geometrical phenomena, the existence of which could be explained in many ways.

At the beginning of the 1970s, his life had the most dramatic changes, 
as he had to make use of a wheelchair permanently, 
and lost completely the ability to handwrite [2].  
Seemingly, his neurons, the source of his intellect, were disappearing. 
This coincides with his 1974 Nature paper [7] on black hole evaporation. 

ALS results in difficulty when speaking, swallowing, and eventually breathing. 
It is a well-known fact that, in the second half of the 1970s, Stephen Hawking developed serious problems communicating with his relatives. 
It is only natural that he worried about just how he was supposed to share his work. 
All that knowledge should not be lost. This is the same lustrum when the paper 
on the breakdown of predictability in gravitational collapse was published [8].  
This paper that was the foundation for the information paradox. 
The paradox of his life. 

We think these points are important because, unlike other realms of human experience, 
physics is often presented as atemporal and disembodied---independent from 
the bodies of those persons doing the physics. 
We challenge this view by complicating the understanding of the development of ideas 
in physics.

One might say that Stephen Hawking ``lost objectivity'' when approaching the black hole puzzle.  This loss of objectivity, in the form of the introduction of fresh metaphors 
(which get into black hole statistics not through the 
traditional thermodynamics approach but through information theory), 
is very welcome and underscores the crucial role of diversity perspective in science.  
Stephen Hawking's disability helped him see the universe under a different light.

\hfil 

\hfil

\noindent 
[1] Dreifus, C. (2011). Life and the cosmos, word by painstaking word. New York Times, May, 10, D1.

\hfil

\noindent 
[2] Ferguson, K. (2011). Stephen Hawking: His life and work. Random House.

\hfil 

\noindent 
[3] Ortega-Rodr\1guez, M., Sol\1s-S\'anchez, H., Boza-Oviedo, E., Chaves-Cruz, K., 
Guevara-Bertsch, M., Quir\'os-Rojas, M., Vargas-Hern\'andez, S. \& Venegas-Li, A. (2017). 
The Early Scientific Contributions of J. Robert Oppenheimer: Why Did the Scientific Community Miss the Black Hole Opportunity? Physics in Perspective, 19(1), 60--75.

\hfil 

\noindent  
[4] Hawking, S. W. \& Sciama, D. W. (1969). Singularities in Collapsing Stars and Expanding Universes. Comments on Astrophysics and Space Physics, 1(1).

\hfil 

\noindent  
[5] Gribbin, J. \& White, M. (2016). Stephen Hawking: a life in science. Pegasus Books.

\hfil 

\noindent  
[6] Penrose, R. (1969). Gravitational collapse: The role of general relativity. Riv. Nuovo Cim., 1, 1141--1165.

\hfil 

\noindent 
[7] Hawking, S. W. (1974). Black hole explosions? Nature, 248(5443), 30.

\hfil 

\noindent  
[8] Hawking, S. W. (1976). Breakdown of predictability in gravitational collapse. Physical Review D, 14(10), 2460.

\end{document}